# Field Equations and Conservation Laws Derived from the Generalized Einstein's Lagrangian Density for A Gravitational System and Their Implications to Cosmology

**Fang-Pei Chen**
Department of Physics, Dalian University of Technology, Dalian 116024, China.
E-mail: chenfap@dlut.edu.cn

**Abstract**   Based on investigations of the fundamental properties of the generalized Einstein's Lagrangian density for a gravitational system, the theoretical foundations of the modified Einstein's field equations and the Lorentz and Levi-Civita's conservation laws are systematically studied. The theory of cosmology developed on the basis of these equations and laws is analyzed in detail. Some new properties and new effects of the cosmos are deduced; these new properties and new effects could be tested via future experiments and observations.

## 1. Introduction

In gravitational theories the action integral $I = \int \sqrt{-g(x)} L(x) d^4 x$ is always used to study the laws of gravitation [1,2,3]. Where g(x) is the determinant of $g_{\mu\nu}(x)$; $\sqrt{-g(x)} L(x)$ is the total Lagrangian density of a gravitational system, which can be divided into two parts: $\sqrt{-g(x)} L(x) = \sqrt{-g(x)} L_M(x) + \sqrt{-g(x)} L_G(x)$. Here $\sqrt{-g(x)} L_G(x)$ is called gravitational Lagrangian density which is composed of gravitational fields only, $\sqrt{-g(x)} L_M(x)$ is called matter Lagrangian density which is composed of both matter fields and gravitational fields [1]. Therefore $\sqrt{-(g)} L_G(x)$ describes only pure gravitational fields; but apart from describing matter fields, $\sqrt{-g(x)} L_M(x)$ also describes the interactions between gravitational field and matter field.

The gravitational field equations can be derived from $\sqrt{-g(x)} L_M(x) + \sqrt{-g(x)} L_G(X)$. It is well known, for the theory of General Relativity, that the gravitational field equations are

$$R^{\mu\nu} - \frac{1}{2} g^{\mu\nu} R - \lambda g^{\mu\nu} = -8\pi G T^{\mu\nu}_{(M)} \qquad (1)$$

where $\lambda$ is cosmological constant, whose value may be equal to zero or nonzero. Eq.(1) can be derived from [4]

$$L_M(x) = L_M[\psi(x); \psi_{|\mu}(x); h^i_{.\mu}(x)]$$

and
$$L_G(x) = \frac{1}{16\pi G}[R(x) + 2\lambda] \qquad (2)$$

where $\psi(x)$ is the matter field, $\psi_{|\mu}(x)$ is the covariant derivative of $\psi(x)$:

$$\psi_{|\mu}(x) = \psi_{,\mu}(x) + \frac{1}{2} h_i^\lambda(x) h_{j\lambda,\mu}(x) \sigma^{ij} \psi(x) \quad [4] \qquad (3)$$

We shall call $\sqrt{-g(x)} L_M(x)$ and $\sqrt{-(g)} L_G(x)$ in Eq.(2) the Einstein's Lagrangian density.

Since the great majority of the fundamental fields for matter field are spinors, it is necessary to use tetrad field $h^i_{.\mu}(x)$ [4]. The metric field $g_{\mu\nu}(x)$ is expressed as $g_{\mu\nu}(x) = h^i_{.\mu}(x) h^j_{.\nu}(x) \eta_{ij}$ and we have

$$h_i^{.\mu}(x) = \eta_{ij} g_{\mu\nu}(x) h^j_{.\nu}(x) \quad ; \qquad h_{i\nu,\lambda}(x) = \frac{\partial}{\partial x^\lambda} h_{i\nu}(x) \quad ; \qquad \text{etc.}$$

Based on Eq.(3) $\sqrt{-g(x)} L_M(x)$ can be denoted by the following functional form:

$$\sqrt{-g(x)} L_M(x) = \sqrt{-g(x)} L_M[\psi(x); \psi_{,\lambda}(x); h^i_{.\mu}(x); h^i_{.\mu,\lambda}(x)] \qquad (4)$$

Since
$$R = g^{\mu\nu}\left(\frac{\partial \Gamma^\lambda_{..\mu\nu}}{\partial x^\lambda} - \frac{\partial \Gamma^\lambda_{..\mu\lambda}}{\partial x^\nu} + \Gamma^\sigma_{..\mu\nu}\Gamma^\lambda_{..\lambda\sigma} - \Gamma^\sigma_{..\mu\lambda}\Gamma^\lambda_{..\nu\sigma}\right),$$

where $\Gamma^\lambda_{..\mu\nu} = \frac{1}{2} g^{\lambda\sigma}(g_{\sigma\mu,\nu} + g_{\sigma\nu,\mu} - g_{\mu\nu,\sigma})$ is the Christoffel symbol. Because $\lambda$ and other parameters are only nondynamical constant or function, so $\sqrt{-g(x)} L_G(x)$ in Eq.(2) can be denoted by the functional form of the dynamical fields $h^i_{.\mu}(x)$ and their derivatives:

$$\sqrt{-g(x)} L_G(x) = \sqrt{-g(x)} L_G[h^i_{.\mu}(x); h^i_{.\mu,\lambda}(x); h^i_{.\mu,\lambda\sigma}(x)] \qquad (5)$$

Eq.(5) shows that $h^i_{.\mu}(x)$ are the dynamical gravitational fields.

Eq.(4) and Eq.(5) are summarized from the general character of General Relativity, but they have much wider ranges of application. For example, Eq.(4) and Eq.(5) also describe the general character of the gravitational theory for those systems whose Lagranges are

$$L_M(x) = L_M [\psi(x); \psi_{|\mu}(x); h^i_{\cdot\mu}(x)]$$

$$L_G(x) = \frac{1}{16\pi G}[R(x) + 2\lambda + 2D(x)] \quad (6)$$

In Eq.(6), $D(x)$ is some nondynamical scalar function of $\{x\}$. So we will call $\sqrt{-g(x)} L_M(x)$ and $\sqrt{-(g)} L_G(x)$ denoted by Eq.(4) and Eq.(5) the generalized Einstein's Lagrangian density for a gravitational system.

Many gravitational theories within the applicable ranges of Eq.(4) and Eq.(5) must all have the general properties deduced from them which will be described below. Especially we will show that the Lorentz and Levi-Civita's conservation laws

$$\frac{\partial}{\partial x^\mu}(\sqrt{-g} T^{\mu\nu}_{(M)} + \sqrt{-g} T^{\mu\nu}_{(G)}) = 0 \quad (7)$$

can be derived from Eq.(4) and Eq.(5); these conservation laws are correct and rational and suitable for many gravitational theories including General Relativity.

On the other hand, gravitational field equations derived from different Lagrangian densities must be different. From Lagranges expressed in Eq.(6) the modified Einstein field equations

$$R^{\mu\nu} - \frac{1}{2} g^{\mu\nu} R - \lambda g^{\mu\nu} - D^{\mu\nu} = -8\pi G T^{\mu\nu}_{(M)} \quad (8)$$

are derived (where $D^{\mu\nu} = D g^{\mu\nu}$) which are different from the Einstein field equations in mathematical form and in physical meaning.

In this work we will study the cosmology under the influences of the modified Einstein's field equations Eq.(8) and the Lorentz and Levi-Civita's conservation laws Eq.(7). This theory of cosmology will lead to the following distinct properties and effects of cosmos: the energy of matter field might originate from the gravitational field; the big bang might not have occurred; the fields of the dark energy and some parts of the dark matter would not be matter fields but might be gravitational fields. These distinct properties and effects will be discussed in detail below.

**2. The fundamental properties of the generalized Einstein's Lagrangian density for a gravitational system**

Symmetries exist universally in physical systems, one fundamental symmetry of a gravitational system is that the action integrals

$$I_M = \int \sqrt{-g(x)} L_M(x) d^4x \qquad I_G = \int \sqrt{-g(x)} L_G(x) d^4x \qquad \text{and}$$

$$I = I_M + I_G = \int \sqrt{-g(x)} (L_M(x) + L_G(x)) d^4x$$

satisfy $\delta I_M = 0$, $\delta I_G = 0$ and $\delta I = 0$ respectively under the following two simultaneous transformations [1,5]:

(1), the infinitesimal general coordinate transformation

$$x^\mu \to x'^\mu = x^\mu + \xi^\mu(x) \tag{9}$$

(2), the local Lorentz transformation of tetrad frame

$$e_i(x) \to e'_i(x') = e_i(x) - \varepsilon^{mn}(x)\delta_m^j \eta_{ni} e_j(x) \tag{10}$$

The symmetry (1) is precisely the symmetry of local space-time translations.

The sufficient condition of an action integral $I = \int \sqrt{-g(x)}\, L(x)\, d^4x$ being $\delta I = 0$ under above transformations is [1,6]:

$$\delta_0(\sqrt{-g}L) + (\xi^\mu \sqrt{-g}L)_{,\mu} \equiv 0 \tag{11}$$

where $\delta_0$ represent the variation at a fixed value of $x$. Evidently there are also the relations

$$\delta_0(\sqrt{-g}L_M) + (\xi^\mu \sqrt{-g}L_M)_{,\mu} \equiv 0 \ ; \ \delta_0(\sqrt{-g}L_G) + (\xi^\mu \sqrt{-g}L_G)_{,\mu} \equiv 0 \tag{12}$$

If there exists only the symmetry (2), Eqs.(11,12) reduce to $\delta_0(\sqrt{-g}L) \equiv 0$, $\delta_0(\sqrt{-g}L_M) \equiv 0$ and $\delta_0(\sqrt{-g}L_G) \equiv 0$ respectively.

From Eq.(4) and Eq.(5) we have

$$\delta_0(\sqrt{-g}L_M) = \frac{\partial(\sqrt{-g}L_M)}{\partial \psi}\delta_0\psi + \frac{\partial(\sqrt{-g}L_M)}{\partial \psi_{,\lambda}}\delta_0\psi_{,\lambda} + \frac{\partial(\sqrt{-g}L_M)}{\partial h^i_{\cdot\mu}}\delta_0 h^i_{\cdot\mu}$$
$$+ \frac{\partial(\sqrt{-g}L_M)}{\partial h^i_{\cdot\mu,\lambda}}\delta_0 h^i_{\cdot\mu,\lambda} \tag{13}$$

$$\delta_0(\sqrt{-g}L_G) = \frac{\partial(\sqrt{-g}L_G)}{\partial h^i_{\cdot\mu}}\delta_0 h^i_{\cdot\mu} + \frac{\partial(\sqrt{-g}L_G)}{\partial h^i_{\cdot\mu,\lambda}}\delta_0 h^i_{\cdot\mu,\lambda} + \frac{\partial(\sqrt{-g}L_G)}{\partial h^i_{\cdot\mu,\lambda\sigma}}\delta_0 h^i_{\cdot\mu,\lambda\sigma} \tag{14}$$

As $\psi(x)$ is spinor and $h^i_{\cdot\mu}(x)$ is both tetrad Lorentz vector and coordinate vector, under the infinitesimal general coordinate transformation and the local Lorentz transformation of tetrad frame, it is not difficult to derive the following induced variations [3]:

$$\delta_0 \psi(x) = \frac{1}{2} \varepsilon^{mn}(x) \sigma_{mn} \psi(x) - \xi^\alpha(x) \psi_{,\alpha}(x) \tag{15}$$

$$\delta_0 \psi_{,\lambda}(x) = \frac{1}{2} \varepsilon^{mn}(x) \sigma_{mn} \psi_{,\lambda}(x) - \frac{1}{2} \varepsilon^{mn}_{,\lambda}(x) \sigma_{mn} \psi(x) - \xi^\alpha(x) \psi_{,\alpha\lambda}(x) - \xi^\alpha_{,\lambda}(x) \psi_{,\alpha}(x) \tag{16}$$

$$\delta_0 h^i_{\cdot\mu}(x) = \varepsilon^{mn}(x) \delta^i_m \eta_{nj} h^j_{\cdot\mu}(x) - \xi^\alpha_{,\mu}(x) h^i_{\cdot\alpha}(x) - \xi^\alpha(x) h^i_{\cdot\mu,\alpha}(x) \tag{17}$$

$$\delta_0 h^i_{\cdot\mu,\lambda}(x) = \varepsilon^{mn}(x) \delta^i_m \eta_{nj} h^j_{\cdot\mu,\lambda}(x) + \varepsilon^{mn}_{,\lambda}(x) \delta^i_m \eta_{nj} h^j_{\cdot\mu}(x) - \xi^\alpha_{,\mu}(x) h^i_{\cdot\alpha,\lambda}(x)$$
$$- \xi^\alpha_{,\mu\lambda}(x) h^i_{\cdot\alpha}(x) - \xi^\alpha(x) h^i_{\cdot\mu,\alpha\lambda}(x) - \xi^\alpha_{,\lambda}(x) h^i_{\cdot\mu,\alpha} \tag{18}$$

$$\delta_0 h^i_{\cdot\mu,\lambda\sigma}(x) = \varepsilon^{mn}(x) \delta^i_m \eta_{nj} h^j_{\cdot\mu,\lambda\sigma}(x) + \varepsilon^{mn}_{,\sigma} \delta^i_m \eta_{nj} h^j_{\cdot\mu,\lambda}(x)$$
$$+ \varepsilon^{mn}_{,\lambda} \delta^i_m \eta_{nj} h^j_{\cdot\mu,\sigma}(x) + \varepsilon^{mn}_{,\lambda\sigma} \delta^i_m \eta_{nj} h^j_{\cdot\mu}(x) - \xi^\alpha_{,\mu}(x) h^i_{\cdot\alpha,\lambda\sigma}(x)$$
$$- \xi^\alpha_{,\mu\sigma}(x) h^i_{\cdot\alpha,\lambda}(x) - \xi^\alpha_{,\mu\lambda}(x) h^i_{\cdot\alpha,\sigma}(x) - \xi^\alpha_{,\mu\lambda\sigma}(x) h^i_{\cdot\alpha}(x) - \xi^\alpha(x) h^i_{\cdot\mu,\alpha\lambda\sigma}(x)$$
$$- \xi^\alpha_{,\sigma}(x) h^i_{\cdot\mu,\alpha\lambda}(x) - \xi^\alpha_{,\lambda}(x) h^i_{\cdot\mu,\alpha\sigma}(x) - \xi^\alpha_{,\lambda\sigma}(x) h^i_{\cdot\mu,\alpha}(x) \tag{19}$$

Substituting Eqs.(15-19) into Eq.(13) and Eq.(14); using

$$\delta_0(\sqrt{-g}\,\Lambda) + (\xi^\mu \sqrt{-g}\,\Lambda)_{,\mu} \equiv 0,$$ where $\Lambda = L_M$ or $\Lambda = L_G$ or $\Lambda = L_M + L_G$ ;

owing to the independent arbitrariness of $\varepsilon^{mn}(x)$, $\varepsilon^{mn}_{,\lambda}(x)$, $\varepsilon^{mn}_{,\lambda\sigma}(x)$, $\xi^\alpha(x)$, $\xi^\alpha_{,\mu}(x)$, $\xi^\alpha_{,\mu\lambda}(x)$ and $\xi^\alpha_{,\mu\lambda\sigma}(x)$, we obtain the following identities ( if $\Lambda = L_G$, $\dfrac{\partial(\sqrt{-g}\,\Lambda)}{\partial \psi} = 0$, $\dfrac{\partial(\sqrt{-g}\,\Lambda)}{\partial \psi_{,\lambda}} = 0$; if $\Lambda = L_M$, $\dfrac{\partial(\sqrt{-g}\,\Lambda)}{\partial h^i_{\cdot\mu,\lambda\sigma}} = 0$ ):

$$\frac{1}{2}\frac{\partial(\sqrt{-g}\,\Lambda)}{\partial\psi}\sigma_{mn}\psi + \frac{1}{2}\frac{\partial(\sqrt{-g}\,\Lambda)}{\partial\psi_{,\lambda}}\sigma_{mn}\psi_{,\lambda} + \frac{\partial(\sqrt{-g}\,\Lambda)}{\partial h^{m}_{.\mu}}h_{n\mu}$$

$$+\frac{\partial(\sqrt{-g}\,\Lambda)}{\partial h^{m}_{.\mu,\lambda}}h_{n\mu,\lambda} + \frac{\partial(\sqrt{-g}\,\Lambda)}{\partial h^{m}_{.\mu,\lambda\sigma}}h_{n\mu,\lambda\sigma} = 0 \qquad (20)$$

$$\frac{1}{2}\frac{\partial(\sqrt{-g}\,\Lambda)}{\partial\psi_{,\lambda}}\sigma_{mn}\psi + \frac{\partial(\sqrt{-g}\,\Lambda)}{\partial h^{m}_{.\mu,\lambda}}h_{n\mu} + 2\frac{\partial(\sqrt{-g}\,\Lambda)}{\partial h^{m}_{.\mu,\lambda\sigma}}h_{.n\mu,\sigma} = 0 \qquad (21)$$

$$\frac{\partial(\sqrt{-g}\,\Lambda)}{\partial h^{m}_{.\mu,\lambda\sigma}}h_{n\mu} = \frac{\partial(\sqrt{-g}\,\Lambda)}{\partial h^{n}_{.\mu,\lambda\sigma}}h_{m\mu} \qquad (22)$$

$$\frac{\partial(\sqrt{-g}\,\Lambda)}{\partial\psi}\psi_{,\alpha} + \frac{\partial(\sqrt{-g}\,\Lambda)}{\partial\psi_{,\lambda}}\psi_{,\lambda\alpha} + \frac{\partial(\sqrt{-g}\,\Lambda)}{\partial h^{i}_{.\mu}}h^{i}_{.\mu,\alpha}$$

$$+\frac{\partial(\sqrt{-g}\,\Lambda)}{\partial h^{i}_{.\mu,\lambda}}h^{i}_{.\mu,\lambda\alpha} + \frac{\partial(\sqrt{-g}\,\Lambda)}{\partial h^{i}_{.\mu,\lambda\sigma}}h^{i}_{.\mu,\lambda\sigma\alpha} - (\sqrt{-g}\,\Lambda)_{,\alpha} = 0 \qquad (23)$$

$$\frac{\partial(\sqrt{-g}\,\Lambda)}{\partial\psi_{,\lambda}}\psi_{,\alpha} + \frac{\partial(\sqrt{-g}\,\Lambda)}{\partial h^{i}_{.\lambda}}h^{i}_{.\alpha} + \frac{\partial(\sqrt{-g}\,\Lambda)}{\partial h^{i}_{.\mu,\lambda}}h^{i}_{.\mu,\alpha} + \frac{\partial(\sqrt{-g}\,\Lambda)}{\partial h^{i}_{.\lambda,\mu}}h^{i}_{.\alpha,\mu}$$

$$+\frac{\partial(\sqrt{-g}\,\Lambda)}{\partial h^{i}_{.\lambda,\mu\sigma}}h^{i}_{.\alpha,\mu\sigma} + 2\frac{\partial(\sqrt{-g}\,\Lambda)}{\partial h^{i}_{.\mu,\lambda\sigma}}h^{i}_{.\mu,\sigma\alpha} - \sqrt{-g}\,\Lambda\,\delta^{\lambda}_{\alpha} = 0 \qquad (24)$$

$$\frac{\partial(\sqrt{-g}\,\Lambda)}{\partial h^i_{.\alpha,\mu,\lambda}}h^i_{.\alpha} + \frac{\partial(\sqrt{-g}\,\Lambda)}{\partial h^i_{.\mu,\lambda\sigma}}h^i_{.\alpha,\sigma} + \frac{\partial(\sqrt{-g}\,\Lambda)}{\partial h^i_{.\sigma,\lambda\mu}}h^i_{.\sigma,\alpha} - \frac{\partial}{\partial x^\sigma}\left(\frac{\partial(\sqrt{-g}\,\Lambda)}{\partial h^i_{.\mu,\lambda\sigma}}\right)h^i_{.\alpha}$$

$$= -\frac{\partial}{\partial x^\sigma}\left(\frac{\partial(\sqrt{-g}\,\Lambda)}{\partial h^i_{.\mu,\lambda\sigma}}h^i_{.\alpha}\right) \tag{25}$$

$$\frac{\partial(\sqrt{-g}\,\Lambda)}{\partial h^i_{.\mu,\lambda\sigma}}h^i_{.\alpha} + \frac{\partial(\sqrt{-g}\,\Lambda)}{\partial h^i_{.\lambda,\sigma\mu}}h^i_{.\alpha} + \frac{\partial(\sqrt{-g}\,\Lambda)}{\partial h^i_{.\sigma,\mu\lambda}}h^i_{.\alpha} = 0 \tag{26}$$

From Eq.(26) another identity:

$$\frac{\partial^3}{\partial x^\mu \partial x^\lambda \partial x^\sigma}\left(\frac{\partial(\sqrt{-g}\,\Lambda)}{\partial h^i_{.\mu,\lambda\sigma}}h^i_{.\alpha}\right) = 0 \tag{27}$$

can be deduced.

Eqs.(23-26) stem from the symmetry of transformation Eq.(9), and the conservation laws of energy-momentum tensor density for a gravitational system can be derived from these identities; we will discuss the derivation in the subsequent section. Eqs.(20-22) stem from the symmetry of transformation Eq.(10), and the conservation laws of spin density for a gravitational system [1] can be derived from these identities; since the conservation laws of spin density for a gravitational system require studying specially, we shall not discuss the problem about spin density in this paper.

## 3. Equations of fields and Conservation laws of energy-momentum tensor density derived from the generalized Einstein's Lagrangian density

The equations of fields for a gravitational system can be derived from

$$\delta I = \delta \int (\sqrt{-g(x)}\,L(x))\,d^4x = 0 \tag{28}$$

where $\sqrt{-g(x)}\,L(x) = \sqrt{-g(x)}\,L_M(x) + \sqrt{-g(x)}\,L_G(x)$, Eq.(28) can be rewritten into

$$\delta I = \int[\delta_0(\sqrt{-g}\,L) + (\xi^\alpha L)_{,\alpha}]\,d^4x = 0$$, using Gauss' theorem, and setting $\xi^\alpha$ equal to zero

at the integration limits, we get

$$\delta I = \int \delta_0(\sqrt{-g(x)}\,L(x))\,d^4x = 0 \tag{28a}$$

$\delta_0(\sqrt{-g}\,L)$ is the variation of $\sqrt{-g}\,L$ corresponding to the variations of the dynamical field variable for the gravitational system at a fixed value of $x$. Eq.(28a) is equivalent to

$$\delta_0 I = \int \delta_0(\sqrt{-g(x)}\,L(x))\,d^4x = 0 \text{ in form.}$$

If the dynamical field variables of the gravitational system are $\psi(x), h^i_{.\mu}(x)$, then from the generalized Einstein's Lagrangian density *i.e.* Eq.(4) and Eq.(5)

$$\delta_0(\sqrt{-g}\,L) = \frac{\partial(\sqrt{-g}\,L_M)}{\partial \psi}\delta_0\psi + \frac{\partial(\sqrt{-g}\,L_M)}{\partial \psi_{,\lambda}}\delta_0\psi_{,\lambda} + \frac{\partial(\sqrt{-g}\,L)}{\partial h^i_{.\mu}}\delta_0 h^i_{.\mu}$$

$$+ \frac{\partial(\sqrt{-g}\,L)}{\partial h^i_{.\mu,\lambda}}\delta_0 h^i_{.\mu,\lambda} + \frac{\partial(\sqrt{-g}\,L_G)}{\partial h^i_{.\mu,\lambda\sigma}}\delta_0 h^i_{.\mu,\lambda\sigma} = (\frac{\partial(\sqrt{-g}\,L_M)}{\partial \psi} - \frac{\partial}{\partial x^\lambda}\frac{\partial(\sqrt{-g}\,L_M)}{\partial \psi_{,\lambda}})\delta_0\psi$$

$$+ (\frac{\partial(\sqrt{-g}\,L)}{\partial h^i_{.\mu}} - \frac{\partial}{\partial x^\lambda}\frac{\partial(\sqrt{-g}\,L)}{\partial h^i_{.\mu,\lambda}} + \frac{\partial^2}{\partial x^\lambda \partial x^\sigma}\frac{\partial(\sqrt{-g}\,L_G)}{\partial h^i_{.\mu,\lambda\sigma}})\delta_0 h^i_{.\mu}$$

$$+ \frac{\partial}{\partial x^\lambda}(\frac{\partial(\sqrt{-g}\,L_M)}{\partial \psi_{,\lambda}}\delta_0\psi + \frac{\partial(\sqrt{-g}\,L)}{\partial h^i_{.\mu,\lambda}}\delta_0 h^i_{.\mu} + \frac{\partial(\sqrt{-g}\,L_G)}{\partial h^i_{.\mu,\lambda\sigma}}\delta_0 h^i_{.\mu,\sigma}$$

$$- \frac{\partial}{\partial x^\sigma}(\frac{\partial(\sqrt{-g}\,L_G)}{\partial h^i_{.\mu,\lambda\sigma}})\delta_0 h^i_{.\mu,}) \tag{29}$$

where $\delta_0\psi(x), \delta_0 h^i_{.\mu}(x)$ are arbitrary and independent variations, they may be or may not be symmetrical variations.

Substituting Eq.(29) into Eq.(28a), using Gauss' theorem, and setting $\delta_0\psi(x), \delta_0 h^i_{.\mu}(x)$ and their derivatives all equal to zero at the integration limits, we find

$$(\frac{\partial(\sqrt{-g}\,L_M)}{\partial \psi} - \frac{\partial}{\partial x^\lambda}\frac{\partial(\sqrt{-g}\,L_M)}{\partial \psi_{,\lambda}})\delta_0\psi$$

$$+ (\frac{\partial(\sqrt{-g}\,L)}{\partial h^i_{.\mu}} - \frac{\partial}{\partial x^\lambda}\frac{\partial(\sqrt{-g}\,L)}{\partial h^i_{.\mu,\lambda}} + \frac{\partial^2}{\partial x^\lambda \partial x^\sigma}\frac{\partial(\sqrt{-g}\,L_G)}{\partial h^i_{.\mu,\lambda\sigma}})\delta_0 h^i_{.\mu} = 0 \tag{30}$$

Since $\psi(x), h^i_{.\mu}(x)$ are independent dynamical field variables, Eq. (30) is equivalent to the following two equations:

$$\frac{\partial(\sqrt{-g}\,L_M)}{\partial\psi} - \frac{\partial}{\partial x^\mu}\frac{\partial(\sqrt{-g}\,L_M)}{\partial\psi_{,\mu}} = 0 \qquad (31)$$

$$\frac{\partial(\sqrt{-g}\,L_G)}{\partial h^i_{\cdot\mu}} - \frac{\partial}{\partial x^\lambda}\frac{\partial(\sqrt{-g}\,L_G)}{\partial h^i_{\cdot\mu,\lambda}} + \frac{\partial^2}{\partial x^\lambda \partial x^\sigma}\frac{\partial(\sqrt{-g}\,L_G)}{\partial h^i_{\cdot\mu,\lambda\sigma}}$$

$$= -\frac{\partial(\sqrt{-g}\,L_M)}{\partial h^i_{\cdot\mu}} + \frac{\partial}{\partial x^\lambda}\frac{\partial(\sqrt{-g}\,L_M)}{\partial h^i_{\cdot\mu,\lambda}} \qquad (32)$$

Eq.(31) is the equation of matter field; Eq.(32) is the equations of vierbein field $h^i_{\cdot\mu}(x)$ which are gravitational fields.

It is well known that, in the special relativity, the conservation laws of energy-momentum tensor density for a physical system is originated from the action integral $I = \int \sqrt{-g(x)}\,L(x)d^4x$ of this physical system being invariant under space-time finite translations [7]. In relativistic theories of gravitation, there is no symmetry of space-time finite translations but have only the symmetry of local space-time translations $x^\mu \to x'^\mu = x^\mu + \xi^\mu(x)$, which is equivalent to the infinitesimal general coordinate transformation Eq.(9). In the following we will use this local symmetry to deduce some identities which might be regarded as the conservation laws of energy-momentum for a gravitational system.

Eq.(23) can be transformed into

$$\frac{\partial}{\partial x^\lambda}(\sqrt{-g}\,\Lambda\,\delta^\lambda_\alpha - \frac{\partial(\sqrt{-g}\,\Lambda)}{\partial\psi_{,\lambda}}\psi_{,\alpha} - \frac{\partial(\sqrt{-g}\,\Lambda)}{\partial h^i_{\cdot\mu,\lambda}}h^i_{\cdot\mu,\alpha} - \frac{\partial(\sqrt{-g}\,\Lambda)}{\partial h^i_{\cdot\mu,\lambda\sigma}}h^i_{\cdot\mu,\sigma\alpha}$$

$$+ \frac{\partial}{\partial x^\sigma}(\frac{\partial(\sqrt{-g}\,\Lambda)}{\partial h^i_{\cdot\mu,\lambda\sigma}})h^i_{\cdot\mu,\alpha}) = (\frac{\partial(\sqrt{-g}\,\Lambda)}{\partial\psi} - \frac{\partial}{\partial x^\lambda}(\frac{\partial(\sqrt{-g}\,\Lambda)}{\partial\psi_{,\lambda}}))\psi_{,\alpha} \qquad (33)$$

$$+ (\frac{\partial(\sqrt{-g}\,\Lambda)}{\partial h^i_{\cdot\mu}} - \frac{\partial}{\partial x^\lambda}(\frac{\partial(\sqrt{-g}\,\Lambda)}{\partial h^i_{\cdot\mu,\lambda}}) + \frac{\partial^2}{\partial x^\lambda \partial x^\sigma}(\frac{\partial(\sqrt{-g}\,\Lambda)}{\partial h^i_{\cdot\mu,\lambda\sigma}}))h^i_{\cdot\mu,\alpha}$$

Utilizing Eq.(25), Eq.(24) can be transformed into

$$\sqrt{-g}\,\Lambda\,\delta^{\lambda}_{\alpha} - \frac{\partial(\sqrt{-g}\,\Lambda)}{\partial \psi_{,\lambda}}\psi_{,\alpha} - \frac{\partial(\sqrt{-g}\,\Lambda)}{\partial h^{i}_{.\mu,\lambda}}h^{i}_{.\mu,\alpha} - \frac{\partial(\sqrt{-g}\,\Lambda)}{\partial h^{i}_{.\mu,\lambda\sigma}}h^{i}_{.\mu,\sigma\alpha}$$

$$+ \frac{\partial}{\partial x^{\sigma}}(\frac{\partial(\sqrt{-g}\,\Lambda)}{\partial h^{i}_{.\mu,\lambda\sigma}})h^{i}_{.\mu,\alpha} + \frac{\partial^{2}}{\partial x^{\mu}\partial x^{\sigma}}(\frac{\partial(\sqrt{-g}\,\Lambda)}{\partial h^{i}_{.\lambda,\mu\sigma}}h^{i}_{.\alpha}) \qquad (34)$$

$$= (\frac{\partial(\sqrt{-g}\,\Lambda)}{\partial h^{i}_{.\lambda}} - \frac{\partial}{\partial x^{\mu}}(\frac{\partial(\sqrt{-g}\,\Lambda)}{\partial h^{i}_{.\lambda,\mu}}) + \frac{\partial^{2}}{\partial x^{\mu}\partial x^{\sigma}}(\frac{\partial(\sqrt{-g}\,\Lambda)}{\partial h^{i}_{.\lambda,\mu\sigma}}))h^{i}_{.\alpha}$$

Let $\Lambda = L_M + L_G$ and use the equations of fields Eqs.(31,32), from Eq.(33) we get:

$$\frac{\partial}{\partial x^{\lambda}}(\sqrt{-g}\,L_M\,\delta^{\lambda}_{\alpha} - \frac{\partial(\sqrt{-g}\,L_M)}{\partial \psi_{,\lambda}}\psi_{,\alpha} - \frac{\partial(\sqrt{-g}\,L_M)}{\partial h^{i}_{.\mu,\lambda}}h^{i}_{.\mu,\alpha}$$

$$+ \sqrt{-g}\,L_G\,\delta^{\lambda}_{\alpha} - \frac{\partial(\sqrt{-g}\,L_G)}{\partial h^{i}_{.\mu,\lambda}}h^{i}_{.\mu,\alpha} - \frac{\partial(\sqrt{-g}\,L_G)}{\partial h^{i}_{.\mu,\lambda\sigma}}h^{i}_{.\mu,\sigma\alpha} \qquad (35)$$

$$+ \frac{\partial}{\partial x^{\sigma}}(\frac{\partial(\sqrt{-g}\,L_G)}{\partial h^{i}_{.\mu,\lambda\sigma}})h^{i}_{.\mu,\alpha}) = 0$$

Let $\Lambda = L_M + L_G$ and use the equations of fields Eq.(32), from Eq.(34) we get:

$$\sqrt{-g}(L_M + L_G)\delta^{\lambda}_{\alpha} - \frac{\partial(\sqrt{-g}\,L_M)}{\partial \psi_{,\lambda}}\psi_{,\alpha} - \frac{\partial(\sqrt{-g}(L_M + L_G))}{\partial h^{i}_{.\mu,\lambda}}h^{i}_{.\mu,\alpha}$$

$$- \frac{\partial(\sqrt{-g}\,L_G)}{\partial h^{i}_{.\mu,\lambda\sigma}}h^{i}_{.\mu,\sigma\alpha} + \frac{\partial}{\partial x^{\sigma}}(\frac{\partial(\sqrt{-g}\,L_G)}{\partial h^{i}_{.\mu,\lambda\sigma}})h^{i}_{.\mu,\alpha} + \frac{\partial^{2}}{\partial x^{\mu}\partial x^{\sigma}}(\frac{\partial(\sqrt{-g}\,L_G)}{\partial h^{i}_{.\lambda,\mu\sigma}}h^{i}_{.\alpha}) = 0 \qquad (36)$$

After differentiation Eq.(36) become

$$\frac{\partial}{\partial x^\lambda}(\sqrt{-g}\,L_M\,\delta_\alpha^\lambda - \frac{\partial(\sqrt{-g}\,L_M)}{\partial \psi_{,\lambda}}\psi_{,\alpha} - \frac{\partial(\sqrt{-g}\,L_M)}{\partial h^i_{.\mu,\lambda}}h^i_{.\mu,\alpha}$$

$$+ \sqrt{-g}\,L_G\,\delta_\alpha^\lambda - \frac{\partial(\sqrt{-g}\,L_G)}{\partial h^i_{.\mu,\lambda}}h^i_{.\mu,\alpha} - \frac{\partial(\sqrt{-g}\,L_G)}{\partial h^i_{.\mu,\lambda\sigma}}h^i_{.\mu,\sigma\alpha} \quad (37)$$

$$+ \frac{\partial}{\partial x^\sigma}(\frac{\partial(\sqrt{-g}\,L_G)}{\partial h^i_{.\mu,\lambda\sigma}})h^i_{.\mu,\alpha} + \frac{\partial^2}{\partial x^\mu \partial x^\sigma}(\frac{\partial(\sqrt{-g}\,L_G)}{\partial h^i_{.\lambda,\mu\sigma}}h^i_{.\alpha})) = 0$$

Eq.(37) is equivalent to Eq.(35) because from Eq.(27), $\dfrac{\partial^3}{\partial x^\lambda \partial x^\mu \partial x^\sigma}(\dfrac{\partial(\sqrt{-g}\,L_G)}{\partial h^i_{.\lambda,\mu\sigma}}h^i_{.\alpha}) = 0.$

Both Eq.(35) and Eq.(37) could be regarded upon as conservation laws of energy-momentum tensor density for gravitational system. Actually, in the special relativity, $L_M\,\delta_\alpha^\lambda - \dfrac{\partial L_M}{\partial \psi_{,\lambda}}\psi_{,\alpha}$ is the energy-momentum tensor of matter field, then

$$\sqrt{-g}\,T^\lambda_{(M)\alpha} = \sqrt{-g}\,L_M\,\delta_\alpha^\lambda - \frac{\partial(\sqrt{-g}\,L_M)}{\partial \psi_{,\lambda}}\psi_{,\alpha} - \frac{\partial(\sqrt{-g}\,L_M)}{\partial h^i_{.\mu,\lambda}}h^i_{.\mu,\alpha} \quad (38)$$

should be interpreted as the energy-momentum tensor density of matter field, $\dfrac{\partial(\sqrt{-g}\,L_M)}{\partial h^i_{.\mu,\lambda}}h^i_{.\mu,\alpha}$ is the influence of gravitational field. In Eq.(35)

$$\sqrt{-g}\,t^\lambda_{(G)\alpha} = \sqrt{-g}\,L_G\,\delta_\alpha^\lambda - \frac{\partial(\sqrt{-g}\,L_G)}{\partial h^i_{.\mu,\lambda}}h^i_{.\mu,\alpha} - \frac{\partial(\sqrt{-g}\,L_G)}{\partial h^i_{.\mu,\lambda\sigma}}h^i_{.\mu,\sigma\alpha} \quad (39)$$

$$+ \frac{\partial}{\partial x^\sigma}(\frac{\partial(\sqrt{-g}\,L_G)}{\partial h^i_{.\mu,\lambda\sigma}})h^i_{.\mu,\alpha}$$

might be interpreted as the energy-momentum tensor density of pure gravitational field. But In Eq.(37)

$$\sqrt{-g}\,T^{.\lambda}_{(G)\alpha} = \sqrt{-g}\,L_G\,\delta^\lambda_\alpha - \frac{\partial(\sqrt{-g}\,L_G)}{\partial h^i_{.\mu,\lambda}} h^i_{.\mu,\alpha} - \frac{\partial(\sqrt{-g}\,L_G)}{\partial h^i_{.\mu,\lambda\sigma}} h^i_{.\mu,\sigma\alpha} \quad (40)$$

$$+ \frac{\partial}{\partial x^\sigma}\left(\frac{\partial(\sqrt{-g}\,L_G)}{\partial h^i_{.\mu,\lambda\sigma}}\right) h^i_{.\mu,\alpha} + \frac{\partial^2}{\partial x^\mu \partial x^\sigma}\left(\frac{\partial(\sqrt{-g}\,L_G)}{\partial h^i_{.\lambda,\mu\sigma}} h^i_{.\alpha}\right)$$

might be interpreted as the energy-momentum tensor density of pure gravitational field.

Thus there are two conservation laws of energy-momentum tensor density for a gravitational system. From Eqs.(35,38,39) we have:

$$\frac{\partial}{\partial x^\lambda}(\sqrt{-g}\,T^{\lambda}_{(M)\alpha} + \sqrt{-g}\,t^{\lambda}_{(G)\alpha}) = 0 \quad (41)$$

which are just the Einstein's conservation laws [8]. From Eqs.(37,38,40) we have:

$$\frac{\partial}{\partial x^\lambda}(\sqrt{-g}\,T^{\lambda}_{(M)\alpha} + \sqrt{-g}\,T^{\lambda}_{(G)\alpha}) = 0 \quad (42)$$

which are just the Lorentz and Levi-Civita's conservation laws, *i.e.* Eq.(7). Although Eq.(37) is equivalent to Eq.(35) in mathematical sense, they are different in physical sense; which will be discussed below:

Let $\Lambda = L_M$ and $\Lambda = L_G$, from Eq.(34) we get the further relations:

$$\sqrt{-g}\,T^{.\lambda}_{(M)\alpha} \stackrel{def}{=} \sqrt{-g}\,L_M\,\delta^\lambda_\alpha - \frac{\partial(\sqrt{-g}\,L_M)}{\partial \psi_{,\lambda}} \psi_{,\alpha} - \frac{\partial(\sqrt{-g}\,L_M)}{\partial h^i_{.\mu,\lambda}} h^i_{.\mu,\alpha}$$

$$= h^i_{.\alpha}\left(\frac{\partial(\sqrt{-g}\,L_M)}{\partial h^i_{.\lambda}} - \frac{\partial}{\partial x^\mu}\left(\frac{\partial(\sqrt{-g}\,L_M)}{\partial h^i_{.\lambda,\mu}}\right)\right) \quad (43)$$

and

$$\sqrt{-g}\, T^{\cdot \lambda}_{(G)\alpha} = \sqrt{-g}\, L_G\, \delta^\lambda_\alpha - \frac{\partial(\sqrt{-g}\, L_G)}{\partial h^i_{\cdot\mu,\lambda}} h^i_{\cdot\mu,\alpha} - \frac{\partial(\sqrt{-g}\, L_G)}{\partial h^i_{\cdot\mu,\lambda\sigma}} h^i_{\cdot\mu,\sigma\alpha}$$

$$+ \frac{\partial}{\partial x^\sigma}\left(\frac{\partial(\sqrt{-g}\, L_G)}{\partial h^i_{\cdot\mu,\lambda\sigma}}\right) h^i_{\cdot\mu,\alpha} + \frac{\partial^2}{\partial x^\mu \partial x^\sigma}\left(\frac{\partial(\sqrt{-g}\, L_G)}{\partial h^i_{\cdot\lambda,\mu\sigma}} h^i_{\cdot\alpha}\right) \quad (44)$$

$$= h^i_{\cdot\alpha}\left(\frac{\partial(\sqrt{-g}\, L_G)}{\partial h^i_{\cdot\lambda}} - \frac{\partial}{\partial x^\mu}\left(\frac{\partial(\sqrt{-g}\, L_G)}{\partial h^i_{\cdot\lambda,\mu}}\right) + \frac{\partial^2}{\partial x^\mu \partial x^\sigma}\left(\frac{\partial(\sqrt{-g}\, L_G)}{\partial h^i_{\cdot\lambda,\mu\sigma}}\right)\right)$$

From Eqs.(28-30), we know $\dfrac{\partial(\sqrt{-g}\, L_M)}{\partial h^i_{\cdot\lambda}} - \dfrac{\partial}{\partial x^\mu}\left(\dfrac{\partial(\sqrt{-g}\, L_M)}{\partial h^i_{\cdot\lambda,\mu}}\right)$ and

$\dfrac{\partial(\sqrt{-g}\, L_G)}{\partial h^i_{\cdot\lambda}} - \dfrac{\partial}{\partial x^\mu}\left(\dfrac{\partial(\sqrt{-g}\, L_G)}{\partial h^i_{\cdot\lambda,\mu}}\right) + \dfrac{\partial^2}{\partial x^\mu \partial x^\sigma}\left(\dfrac{\partial(\sqrt{-g}\, L_G)}{\partial h^i_{\cdot\lambda,\mu\sigma}}\right)$ are the functional derivatives, hence

$T^{\cdot\lambda}_{(M)\alpha}$ and $T^{\cdot\lambda}_{(G)\alpha}$ all are the tensors. But

$$t^{\lambda}_{(G)\alpha} = \sqrt{-g}\, L_G\, \delta^\lambda_\alpha - \frac{\partial(\sqrt{-g}\, L_G)}{\partial h^i_{\cdot\mu,\lambda}} h^i_{\cdot\mu,\alpha} - \frac{\partial(\sqrt{-g}\, L_G)}{\partial h^i_{\cdot\mu,\lambda\sigma}} h^i_{\cdot\mu,\sigma\alpha} + \frac{\partial}{\partial x^\sigma}\left(\frac{\partial(\sqrt{-g}\, L_G)}{\partial h^i_{\cdot\mu,\lambda\sigma}}\right) h^i_{\cdot\mu,\sigma\alpha}$$

is not tensor. Therefore Eq.(42) is covariant under the symmetric transformations denoted by Eq.(9) and Eq.(10); but Eq.(41) lacks the invariant character required by the principles of general relativity, this is the serious defect of Eq.(41).

Eq.(36) tell us that

$$T^\lambda_{(M)\alpha} + T^\lambda_{(G)\alpha} = 0 \quad\quad \text{or} \quad\quad T^{\lambda\nu}_{(M)} + T^{\lambda\nu}_{(G)} = 0 \quad\quad (45)$$

this is an essential property of the Lorentz and Levi-Civita's conservation laws.

Einstein did not agree with Eq.(45) [9], because he believed that the relation expressed by Eq. (45) should make the energy-momentum of a material system, being $T^{\mu\nu}_{(M)} \neq 0$ in the initial state, to reduce spontaneously to $T^{\mu\nu}_{(M)} \to 0$ in the final state. By using Boltzmann's relation S=$k$ ln $N$, we have shown that this view is incorrect [10]. An important debate was evoked about the definitions of energy-momentum tensor density for gravitational field and the related conservation laws in 1917-1918 [9]; Einstein was on the one side of that debate, his opponents are Levi-Civita and others. This debate had not reached unanimity, but because Einstein enjoyed great prestige among academic circles and many scholars followed him, therefore the interpretation by Eq.(39) and the Einstein's conservation laws Eq.(41) have become the prevalent views now in the gravitational theory. The author holds that, as the Lorentz and Levi-Civita's conservation laws being equivalent to the Einstein's conservation laws in mathematical sense, these two conservation laws are all well worth to consider. Which law is correct on physical side ? This question can answer only by experimental and observational tests. To affirm subjectively a law is not suitable. In the last few years the author have thoroughly studied Lorentz and Levi-Civita's conservation laws and found that these conservation laws have rich physical contents [10-13] which can be tested via experiments or observations. In these respects, the Lorentz and Levi-Civita's conservation laws will undoubtedly be used as one of important theoretical foundations to establish a new cosmology.

Both the gravitational Lagrangian density $L_G(x)$ for the Einstein field equations without cosmological constant *i.e.* $R^{\mu\nu} - \frac{1}{2} g^{\mu\nu} R = -8\pi G T^{\mu\nu}_{(M)}$ and the gravitational Lagrangian density $L_G(x)$ for the modified Einstein field equations *i.e.* $R^{\mu\nu} - \frac{1}{2} g^{\mu\nu} R - \lambda g^{\mu\nu} - D^{\mu\nu} = -8\pi G T^{\mu\nu}_{(M)}$ are the special cases of Eq.(5). The Einstein field equations without cosmological constant are the important theoretical foundations for the current cosmology. It is well worth to study whether the modified Einstein field equations could be used also as the important theoretical foundations to establish a new cosmology. We shall discuss this question in next section.

## 4. The modified Einstein field equations, One possible explanation for dark energy and dark matter

The term $D^{\mu\nu}$ in the modified Einstein field equations

$R^{\mu\nu} - \frac{1}{2} g^{\mu\nu} R - \lambda g^{\mu\nu} - D^{\mu\nu} = -8\pi G T^{\mu\nu}_{(M)}$ was introduced firstly by the steady state cosmology [4], but the physical meaning and the method of introduction for $D^{\mu\nu}$ in this paper are different from the steady state cosmology. It must be emphasized that $\lambda g^{\mu\nu}$ and $D^{\mu\nu}$ in the modified Einstein field equations are similar to $(R^{\mu\nu} - \frac{1}{2} g^{\mu\nu} R)$ in essence, they are all derived from the Lagrange for pure gravitational field $L_G$ and therefore they are all quantities used to represent the pure gravitational field. It must be stressed that $I_G = \int \sqrt{-g} L_G d^4 x$ is invariant only under space-time symmetry, but $I_M = \int \sqrt{-g} L_M d^4 x$ is invariant under both space-time symmetry and internal symmetry; so the gravitational field can be acted only by gravitational force and can not be acted by nongravitational forces, but the matter field can be acted by both gravitational force and nongravitational forces. By virtue of $\lambda g^{\mu\nu}$ and $D^{\mu\nu}$ being quantities used to represent the pure gravitational field, these two parts of gravitational field can not interact with nongravitational forces including electromagnetic force, so they must be 'dark'. Hence it is natural to interpret them as 'dark energy' and 'dark matter'.

Instead of using Einstein field equations, we will use the modified Einstein field equations as the theoretical foundation of cosmology. The universe is assumed still to be spatially homogeneous and isotropic (this assumption is called cosmological principle), so the universe has the Robertson-Walker metric [4]

$$d\tau^2 = -dt^2 + a(t)^2 \left\{ \frac{dr^2}{1-kr^2} + r^2 d\theta^2 + r^2 \sin^2\theta d\phi^2 \right\} \tag{46}$$

and the energy-momentum tensor of the matter field should take the form of ideal fluid [4]

$$T^{\mu\nu}_{(M)} = (\rho_M + p_M) u^\mu u^\nu + p_M g^{\mu\nu} . \tag{47}$$

Using the same method described in chapter 15 of reference [4], from Eqs. (8,46,47) we can derive the following two equations:

$$\left(\frac{da}{dt}\right)^2 + k = \frac{8\pi G}{3}\left(\rho_M + \frac{\lambda}{8\pi G} + \frac{D}{8\pi G}\right) a^2 \tag{48}$$

$$\frac{d^2 a}{dt^2} = -\frac{4\pi G}{3}\left(\rho_M + 3p_M - \frac{\lambda}{4\pi G} - \frac{D}{4\pi G}\right) a \tag{49}$$

where $\rho_M$ is the mean energy density of matter, and $p_M$ is the mean pressure of matter. The cosmological principle demands that $a, \rho_M, p_M, D$ all depend on the cosmic standard time only [4], *i.e.* they all are functions of $t$: $a(t), \rho_M(t), p_M(t), D(t)$. We shall show below that at present time $(\rho_M + 3p_M - \frac{\lambda}{4\pi G} - \frac{D}{4\pi G}) < 0$, hence $\frac{d^2 a}{dt^2} > 0$, *i.e.* the universe is expanding accelerative.

The four quantities $H(t)$, $q(t)$, $\rho_c$, $\Omega_{(M)}$ are used frequently in cosmology, the definitions of $H(t)$, $q(t)$ are: $\quad H(t) = \dfrac{da(t)/dt}{a(t)}, \quad q(t) = -\dfrac{d^2 a(t)}{dt^2} \dfrac{a(t)}{\left(\dfrac{da(t)}{dt}\right)^2};$

when $t_0$ is the present time, the parameters $H_0 = H(t_0)$, $q_0 = q(t_0)$ are known as Hubble's constant and the deacceleration parameter respectively; the definitions of $\rho_c$, $\Omega_{(M)}$ are

$$\rho_c = \dfrac{3 H_0^2}{8\pi G}, \qquad \Omega_{(M)} = \dfrac{\rho_M(t_0)}{\rho_c}$$

which are called critical density and density parameter respectively. From Eqs.(48) and (49) and using these parameters the following relations can be obtained:

$$\dfrac{k}{H_0^2 a^2(t_0)} = (2q_0 - 1) + \dfrac{(\lambda + D(t_0))}{H_0^2} \tag{50}$$

$$\Omega_{(M)} + \dfrac{1}{8\pi G \rho_c}(\lambda + D(t_0)) = 1 \tag{51}$$

To derive Eq. (50) we have used the fact that the matter energy density of the present university is dominated by nonrelativistic matter, so $p_M(t_0) \ll \rho_M(t_0)$ and $p_M(t_0)$ can be neglected.

Eq. (50) implies that when

$$2q_0 = 1 - \dfrac{(\lambda + D(t_0))}{H_0^2} \tag{52}$$

then $k = 0$. It has been determined from astronomical observations [14] that $k$ is close to the value 0; so Eq. (52) must be satisfied. If the two terms $\lambda g^{\mu\nu}$ and $D^{\mu\nu}$ in Eq. (8) do not exist, i.e. $\lambda = 0, D(t) = 0$, Eq.(52) becomes $2q_0 = 1$. If we define $\rho_\lambda = \dfrac{\lambda}{8\pi G}$, $\Omega_{(\lambda)} = \dfrac{\rho_\lambda}{\rho_c}$; $\rho_D = \dfrac{D(t)}{8\pi G}$, $\Omega_{(D)} = \dfrac{\rho_D(t_0)}{\rho_c}$; then Eq. (51) becomes

$$\Omega_{(M)} + \Omega_{(\lambda)} + \Omega_{(D)} = 1 \tag{53}$$

Eq.(53) means that although $\lambda g^{\mu\nu}$ and $D^{\mu\nu}$ are two quantities which represent the pure gravitational field and are not two quantities which represent matter field in essence, but they have the property that they could be regarded as if they are two parts of energy-momentum tensor of the matter fields. In order to explain this specific property we transform Eq. (8) into

$$R^{\mu\nu} - \frac{1}{2} g^{\mu\nu} R = -8\pi G T^{\mu\nu}_{mod} \tag{54}$$

where $T^{\mu\nu}_{mod}$ might be called modified energy-momentum tensor;

$$T^{\mu\nu}_{mod} \equiv T^{\mu\nu}_{(M)} - \frac{\lambda}{8\pi G} g^{\mu\nu} - \frac{D^{\mu\nu}}{8\pi G} \tag{55}$$

$T^{\mu\nu}_{mod}$ could also be written as the perfect-fluid form:

$$T^{\mu\nu}_{mod} = (\rho_{mod} + p_{mod}) u^{\mu} u^{\nu} + p_{mod} g^{\mu\nu} \tag{56}$$

comparing Eq. (56) with Eq. (55) we get

$$\rho_{mod} = \rho_M + \rho_\lambda + \rho_D; \quad \rho_\lambda = \frac{\lambda}{8\pi G}, \quad \rho_D = \frac{D}{8\pi G} \tag{57}$$

$$p_{mod} = p_M + p_\lambda + p_D; \quad p_\lambda = -\frac{\lambda}{8\pi G}, \quad p_D = -\frac{D}{8\pi G} \tag{58}$$

The relations for $\rho_\lambda$ and $\rho_D$ conform to the definitions of $\rho_\lambda$ and $\rho_D$ included in Eq. (53).

The author holds the view that there are two kinds of dark matter: one should be the field of $D^{\mu\nu}$ which energy density is $\rho_D$, and the other might be some material matter [14], such as the neutrino, a weakly interacting massive particle (WIMP) and the massive compact halo objects (MACHOs, including low-luminosity stars and black holes), *etc.;* their energy density are some parts of $\rho_M$. The conclusions from CMB data tell us that [14] the Universe is made up as follows: 73% dark energy, 23% dark matter and 4% ordinary (baryonic) matter. According the above view point we would have: $\rho_\lambda(t_0)/\rho_c = 73\%$, $\rho_M(t_0)/\rho_c > 4\%$, $\rho_D(t_0)/\rho_c < 23\%$, and $\rho_M(t_0)/\rho_c + \rho_D(t_0)/\rho_c = 27\%$. The author holds also that we can distinguish between $\rho_D$ and $\rho_M$; because $\rho_D$ can be only acted by gravitational force and can not be acted by nongravitational forces, but $\rho_M$ can be acted by both gravitational force and nongravitational forces, hence it could be possible to distinguish the two kinds of dark matter. These possibilities might be tested by experiments and observations in future.

It must be pointed out that, for the whole cosmos, $\rho_\lambda, \rho_D, \rho_M$ are all less than the critical density $\rho_c = \frac{3[H(t_0)]^2}{8\pi G} = 1.9 h^2 \times 10^{-29} g/cm^3$ [4]; but for a macroscopic gravitational system, $\rho_M \gg \rho_c$, however $\rho_{\lambda,} \rho_D$ still less than $\rho_c$, then from Eq. (58) $\rho_{mod} \approx \rho_M$, therefore Eq. (8) $R^{\mu\nu} - \frac{1}{2}g^{\mu\nu}R - \lambda g^{\mu\nu} - D^{\mu\nu} = -8\pi G T^{\mu\nu}_{(M)}$ degenerate to. $R^{\mu\nu} - \frac{1}{2}g^{\mu\nu}R = -8\pi G T^{\mu\nu}_{(M)}$.

The Eq.(49) can be rewritten as $\frac{d^2 a}{dt^2} = -\frac{4\pi G}{3}(\rho_M + 3p_M - 2\rho_\lambda - 2\rho_D)a$; owing to $p_M(t_0) \ll \rho_M(t_0)$ and utilizing the above CMB data, it is evidently $(\rho_M + 3p_M - 2\rho_\lambda - 2\rho_D) < 0$, therefore $\frac{d^2 a}{dt^2} > 0$, i.e. the universe is accelerating in its expansion.

It had been suggested by some scholars that the energy density $\rho_\lambda$ of field $-\frac{\lambda}{8\pi G}g_{\mu\nu}$ perhaps might be equal to the vacuum energy density of matter field [15]. However their views all run into some difficulties and conflicting issues. But according to the author's view point $\rho_\lambda$ is a part of gravitational field's energy density which belongs to $T^{\mu\nu}_{(G)}$, but the vacuum energy density of matter field is a part of matter field's energy density $\rho_M$ which belongs to $T^{\mu\nu}_{(M)}$; they might be different in essence, and there is no relation between $\rho_\lambda$ and the vacuum energy density of matter field. As we have indicated above that $\rho_\lambda$ can be only acted by gravitational force and can not be acted by nongravitational forces, but $\rho_M$ can be acted by both gravitational force and nongravitational forces, so it could be possible to distinguish $\rho_\lambda$ from the vacuum energy density of matter field by future experiments and observations.

**5. The Lorentz and Levi-Civita's conservation laws, One possible explanation for the origin of matter field's energy**

From Eq. (45) we get $\triangle T^{\mu\nu}_{(M)} = -\triangle T^{\mu\nu}_{(G)}$ immediately, this relation means that for an isolated gravitational system if the energy-momentum of matter field increases, then the energy-momentum of gravitational field should decrease, *i.e.* the energy-momentum of gravitational field might transform into the energy-momentum of matter field. This possibility might occur in reality, since the number of microscopic states both for matter field and gravitational field should all increase in this process so that the entropy of the system increases. It is worth pointing out that in the above process the absolute value of gravitational field energy is increasing, thus the number of microscopic states for gravitational field should also increase. This possibility could be used to explain the origin of matter field's energy and this explanation is one important consequence of Lorentz and Levi-Civita's conservation laws. Before discussing this problem we will deduce some relations from the Lorentz and Levi-Civita's conservation laws first.

By comparing Eq.(45) with Eq.(8), we get

$$T^{\mu\nu}_{(G)} = \frac{1}{8\pi G}(R^{\mu\nu} - \frac{1}{2}g^{\mu\nu}R - \lambda g^{\mu\nu} - D^{\mu\nu}) \qquad (59)$$

this equality means $T^{\mu\nu}_{(G)}$ can be divided into three parts:

$$T^{\mu\nu}_{(G)} = {}^{R}T_{(G)}\mu\nu + {}^{\lambda}T_{(G)}\mu\nu + {}^{D}T_{(G)}\mu\nu \tag{60}$$

where ${}^{R}T_{(G)}\mu\nu = \frac{1}{8\pi G}(R^{\mu\nu} - \frac{1}{2}g^{\mu\nu}R)$ is the part of gravitational field's energy-momentum due to space-time curvature; ${}^{\lambda}T_{(G)}\mu\nu = -\frac{\lambda g^{\mu\nu}}{8\pi G}$ is the part of gravitational field's energy-momentum due to cosmological constant; ${}^{D}T_{(G)}\mu\nu = -\frac{D^{\mu\nu}}{8\pi G}$ is the part of gravitational field's energy-momentum due to the correction field $D^{\mu\nu}$.

From Eqs. (7,8,45,59,60) we obtain

$$ {}^{R}T_{(G)}\mu\nu + {}^{\lambda}T_{(G)}\mu\nu + {}^{D}T_{(G)}\mu\nu + T_{(M)}{}^{\mu\nu} = 0 \tag{61}$$

$$\frac{\partial}{\partial x^{\mu}}({}^{R}T_{(G)}\mu\nu + {}^{\lambda}T_{(G)}\mu\nu + {}^{D}T_{(G)}\mu\nu + T_{(M)}{}^{\mu\nu}) = 0 \tag{62}$$

Let $\mu = \nu = 0$ in Eqs. (61,62) then we have

$$\rho_R + \rho_\lambda + \rho_D + \rho_M = 0 \tag{63}$$

$$\frac{d}{dt}(\rho_R + \rho_\lambda + \rho_D + \rho_M) = 0 \tag{64}$$

$\rho_G = \rho_R + \rho_\lambda + \rho_D$ is the total energy density of the pure gravitational field, $\rho_R, \rho_\lambda$ or $\rho_D$ is the energy density relating to $T_{(G)}^{R}{}^{\mu\nu}$, $T_{(G)}^{\lambda}{}^{\mu\nu}$ or $T_{(G)}^{D}{}^{\mu\nu}$ respectively.

If we assumes that in all circumstances $\rho_\lambda \geq 0$, $\rho_D \geq 0$ and $\rho_M \geq 0$, then $\rho_R \leq 0$ and $\rho_G = \rho_R + \rho_\lambda + \rho_D \leq 0$ are true in all circumstances as well.

The above results might be applied to cosmology; from Eqs(54-58) and using the same methods of Ref.[4], we can get the relations:

$$\frac{d(\rho_M + \rho_D)}{dt} + 3\frac{\frac{da}{dt}}{a}(\rho_M + \rho_D + p_M + p_D) = 0 \qquad (65)$$

Eq. (65) is equivalent to Eq. (64), there exists the relation:

$$\frac{d(\rho_R + \rho_\lambda)}{dt} = 3\frac{\frac{da}{dt}}{a}(\rho_M + \rho_D + p_M + p_D), \text{ i.e. } \frac{d\rho_R}{dt} = 3\frac{\frac{da}{dt}}{a}(\rho_M + p_M),$$

since Eqs (57, 58) tell us $\rho_D + p_D = 0$, $\rho_\lambda = const$. Besides, $3\frac{\frac{da}{dt}}{a} \cong \frac{\Delta V}{V \Delta t}$, so the Eq.(65) can be rewritten as

$$\frac{\Delta \rho_D}{\Delta t} \cong -\frac{\Delta(\rho_M V)}{V \Delta t} - p_M \frac{\frac{\Delta V}{\Delta t}}{V}, \qquad (66)$$

where V is any volume in the space, $\frac{\Delta \rho_D}{\Delta t}$ represents the rate of energy density change for the field $D^{\mu\nu}$, $\frac{\Delta(\rho_M V)}{V \Delta t} = \frac{\Delta \rho_M}{\Delta t} + \frac{\rho_M \Delta V}{V \Delta t}$ represents the total energy change per unit volume per unit time for the matter field, $p_M \dfrac{\Delta V / \Delta t}{V}$ represents the work done per unit volume per unit time by the matter field. For matter field, there are $\rho_M \geq 0$ and $p_M \geq 0$ usually; if $\Delta V > 0$, $\Delta \rho_M > 0$ or $\Delta \rho_M < 0$ but $\Delta \rho_M + \dfrac{\rho_M \Delta V}{V} > 0$, then $\dfrac{\Delta(\rho_M V)}{V \Delta t} > 0$, *i.e.* the energy of the matter field increases, and $p_M \dfrac{\Delta V}{V \Delta t} > 0$, *i.e.* the work done by the matter field is positive. Hence from Eq. (66) $\dfrac{\Delta \rho_D}{\Delta t} < 0$, this relation means that some energy of field $D^{\mu\nu}$ has transformed into the energy of matter field. This tells us that the increase of matter field energy stems from the decrease of gravitational field energy.

If we assume that at initial time $t=0$, the state of cosmos is $\rho_M = 0$, $p_M = 0$ everywhere, *i.e.* $T^{\mu\nu}{}_{(M)}(0) = 0$ everywhere (Since $T^{\mu\nu}{}_{(M)} = p_M g^{\mu\nu} + (\rho_M + p_M) U^\mu U^\nu$); then according to the above analysis, the energy of matter field would be transformed from the gravitational field continuously, this means that the energy of matter field might originate from the gravitational field.

The state $T^{\mu\nu}{}_{(M)}(0) = 0$ is the lowest state of energy-momentum for the matter field in the universe. It must be emphasized that this state is not equal to the other lower energy state, *i.e.* the so called 'vacuum' state of quantum matter field; since at the 'vacuum' state, $\rho_M > 0$. On the other hand, it must be indicated that the energy creation of matter field does not mean the matter field creation, thus if at $t=0$, $\rho_M(0) = 0$, at $t>0$, $\rho_M(t) > 0$, it means only that the state of matter field changes from the lowest state to a higher state, but the matter field exists all along from the beginning of the energy change. It must be indicated also that the cosmos might have not the state with $T^{\mu\nu}{}_{(M)}(0) = 0$ everywhere in the space, *i.e.* the state $t=0$, $\rho_M = 0$, $p_M = 0$ everywhere does not exist. Why is there not this state? This is due to the quantum fluctuations, at any time there must always be energy-momentum transformations between gravitational field and matter field, so the state $t=0$, $\rho_M = 0$, $p_M = 0$ everywhere is not possible. The standard cosmology (SBBC) has a beginning state called big bang, and it is assumed that the total energy of matter fields (including the inflation field) had existed since the big bang. At the big bang, *i.e.* at $t=0$, it is generally thought that $\rho_M \to \infty$ and the temperature $T \to \infty$. Moreover, SBBC does not study the origin of the matter field's energy. As we have shown in the above discussions, the energy of matter field might be transformed from the gravitational field continuously, and the universe could be expanding without need for the state $\rho_M \to \infty$, this means that the big bang might never have occurred.

How does the energy-momentum transform from the gravitational field into the matter field? How the cosmos evolve? These problems relate to quantum theory of gravitational field. As a complete and consistent quantum theory of gravitational field has not yet been constructed yet, we can not answer this problem clearly and completely. None the less, we could propose the following assumptions which can be proved, or refuted, or revised by future experiments and observations:

(1). The energy of gravitational field might transform into the energy of some elementary particles (including the thermal energy of elementary particles); but these transformed energy can not lead to the state $\rho_M \to \infty$ and the temperature can not reach $T \to \infty$.

(2). In the past, when some conditions were satisfied; some eras, which were similar to the eras of the early universe in SBBC [1], might emerge from the quantum fluctuation. But in the theory of cosmology developed on the basis of the Lorentz and Levi-Civita's conservation laws and the modified Einstein field equations, the period of every era might be longer than that in SBBC. As an example we will use Eq.(66) to show that the cosmic change taken place in the matter field for the radiation-dominated era:

Rewriting Eq.(66) $\dfrac{\Delta \rho_D}{\Delta t} \cong -\dfrac{\Delta(\rho_M V)}{V \Delta t} - p_M \dfrac{\Delta V / \Delta t}{V}$ as

$$\frac{d\rho_D}{dt} = -\frac{d(\rho_M V)}{V dt} - p_M \frac{dV/dt}{V}. \qquad (66')$$

For the radiation-dominated era, $p_M(t) = \frac{1}{3}\rho_M(t)$, Eq.(66') become

$\frac{d\rho_M}{\rho_M} + 4\frac{da}{a} = -d\rho_D$. In SBBC, $\rho_D = 0$, we will get $\rho_M a^4 = 1$; in the theory of cosmology founded on the Lorentz and Levi-Civita's conservation laws and the modified Einstein field equations, if $dV/dt > 0$, $d\rho_M/dt < 0$ but $\frac{d(\rho_M V)}{V dt} > 0$, then $d\rho_D < 0$, we will get $\rho_M a^4 > 1$. It is obvious that, when the universe expands, $\rho_M$ will decrease slower in the new theory of cosmology than in SBBC.

(3). Especially, there had been the change from the radiation-dominated era to the matter-dominated era which is similar with SBBC. At the radiation-dominated era, matter and radiation were presumably in thermal equilibrium; their temperature is higher than $4000\,^oK$. When the temperature is below $4000\,^oK$, the matter-dominated era commenced, and the radiation existed still and had been red-shifted owing to the expansion of the universe. It is widely believed that the microwave radiation background is just the left-over radiation [4] So that the new theory of cosmology can explain the microwave radiation background as well as SBBC.

In SBBC the observed abundances of light nuclei in the universe are explained as the result of nucleon-synthesis taking place in the early universe at a temperature of about $10^9\,^oK$. In the theory of cosmology developed on the basis of the Lorentz and Levi-Civita's conservation laws and the modified Einstein field equation, although the observed abundances of light nuclei in the universe can be explained with the same reason as SBBC, there is another explanation which had put forward by some cosmologists in the 1950's. They had studied the possibilities of that the light nuclei in the universe might be formed from hydrogen nuclei in the interiors of stars [4]; but the cosmic abundance of helium is too large to be easily explained in terms of nucleon-synthesis in the interiors of stars at $10^{10}$ years estimated by SBBC. However in the new theory of cosmology, the period of every era might be longer than SBBC, the helium nuclei in the universe might be synthesized in a longer time frame; therefore this problem does not exist. Which explanation is correct will be determined by future tests.

## 6. Conclusions

In this paper it has been shown that: 1), The modified Einstein field equations are rational as well as the Einstein field equations; the Lorentz and Levi-Civita's conservation laws are equivalent to the Einstein's conservation laws mathematically. Just as SBBC is developed on the basis of the Einstein's conservation laws and the Einstein field equations, it is quite reasonable to establish a new theory of cosmology developed on the basis of the Lorentz and Levi-Civita's conservation laws and the modified Einstein field equations. 2), Some new properties and new effects are deduced from the new theory of cosmology, these new properties and new effects could be tested via future experiments and observations.

As many new evidences of observations [14, 16, 17] have brought out some crucial weaknesses of SBBC. It is necessary to introduce new concepts and new theories, so I believe that it is significant to study the theory of cosmology founded on the Lorentz and Levi-Civita's conservation laws and the modified Einstein field equations.